\documentclass[aps,showpacs,nofootinbib]{revtex4}
\usepackage{amsmath} % need for subequations
\usepackage{amssymb}
\usepackage{graphicx} % need for figures
\usepackage{color} % use if color is used in text
\usepackage{mathtext}
\usepackage{epstopdf}

\begin{document}
\newcommand{\blue}[1]{\textcolor{blue}{#1}}
\newcommand{\new}{\blue}
\newcommand{\green}[1]{\textcolor{green}{#1}}
\newcommand{\modif}{\green}
\newcommand{\red}[1]{\textcolor{red}{#1}}
\newcommand{\attention}{\red}

%Color scheme: {\color{green} green: comments about something along
%the text,}~~{\color{blue} blue: replacing the original text.}
%{\color{red} red: wrong idea or approach.}

\title{Collective effects in strong interaction processes: experimental highlights}

\author{V. A. Okorokov} \email{VAOkorokov@mephi.ru; Okorokov@bnl.gov}
\affiliation{National Research Nuclear University MEPhI (Moscow
Engineering Physics Institute), Kashirskoe highway 31, 115409
Moscow, Russia}

\date{\today}

\begin{abstract}
\noindent {\bf Abstract}---Collective effects are reviewed for
collisions of various systems -- from proton-proton to heavy ion
-- in wide energy range. In proton--proton interactions studies of
hadron jets devote to the better understanding of some basic
features of strong interaction and search for the physics beyond
of Standard Model. First results have been obtained for massive
gauge bosons and antitop-top pair production in proton--nuclear
and heavy ion collisions at multi-TeV energies. The collectivity
has been observed for various particle and beam species, in
particular, in collision of small systems. Experimental results
obtained for discrete symmetries of strong interaction at finite
temperature confirm indirectly the topologically non-trivial
structure of the vacuum. The recent measurements of femtoscopic
correlations provide, in particular, the indirect estimations for
parameters of hyperon-nucleon potentials which are essential for
study of inner structure of compact astrophysical objects. Novel
mechanism for multiparticle production due to collectivity can be
expected in very high energy nuclear collisions and it may be
helpful for better understanding of the nature of the muon puzzle
in ultra-high energy cosmic ray measurements. Thus studies of
collective effects in strong interaction processes provide new
important results for relativistic astrophysics, cosmology and
cosmic ray physics, i.e. have interdisciplinary significance.
\end{abstract}


\maketitle

\section{Introduction}\label{sec:1}

There is no full and self-consistent theory of strong interaction
despite of significant progress during last decades. The general
and well-established picture of strong interaction reaction
\cite{Campbell-book-2018-1} leads to the hypothesis that
collective behavior and corresponding effects are deeply intrinsic
properties of strong interaction driven by the dynamics of the
collision process at all stages of its space-time evolution. As
consequence, investigations of collective effects and collective
modes of excitation of the hadronic / quark-gluon matter provide
unique information and are among the most promising and relevant
for constructing a complete theory of strong interaction and
studying of quantum multiparticle systems. The study of collective
and correlation characteristics of strong interaction makes it
possible to draw conclusions regarding the space-time evolution of
the interaction process and to establish a fundamental
relationship between the geometry and dynamics of the creation of
a final state.

During XXI century the main part of experimental and theoretical
studies within strong interaction physics is based on the research
works making at Relativistic Heavy Ion Collider (RHIC) and the
Large Hadron Collider (LHC). Therefore the consideration below is
focused on the results obtained at these facilities. The
accelerator complex RHIC was designed and was built for
investigations in Quantum chromodynamics (QCD) field specially.
There were 24 successful physics runs since 2000 year. Runs 25 and
26 are planed. The following large detectors are placed at RHIC:
STAR continue to collect new data since 2000, sPHENIX was
commissioned in 2023. Data were also taken by small experiments
BRAHMS, PHOBOS and by large one PHENIX. Table~\ref{tab-1} shows
the data samples collected during 24 runs at RHIC, where
$\sqrt{s_{NN}}$ is the center-of-mass energy per nucleon--nucleon
pair. The sPHENIX and STAR detectors are characterized by good
particle identification and uniform, large acceptance. Thus both
RHIC detectors are suitable rather good for study of various
collective effects. The accelerator complex LHC was designed and
was built for investigations in fundamental physics, in
particular, in subfield of QCD. There were 2 successful physics
runs since 2009 year, the run 3 is in the progress\footnote{It
should be emphasized the run at the LHC implies the multi-year
period of work of the accelerator, namely, run 1 was during
2009--2013 years, run 2 -- 2015--2018 years, run 3 is planned on
2022--2025 years.}. The large detectors ALICE, ATLAS, CMS and LHCb
are placed at the LHC and they collect data since 2009. There also
are wide set of smaller experiments focused on the specific issues
in fundamental physics. The data samples taken at the LHC since
2009 are shown in Table~\ref{tab-2}. ALICE, ATLAS and CMS are a
general-purpose detectors characterized by good particle
identification and (quasi)uniform, large acceptance. Designs of
all of these detectors are well optimized for study of wide sets
of collective effects. Nevertheless the LHCb is single-arm
spectrometer an important results have been already obtained for
femtoscopic correlations and jet physics with this apparatus.

\begin{table}[t!]
\centering \caption{Data samples obtained during RHIC runs.}
\label{tab-1}
\begin{tabular}{ll}\hline
Species & $\sqrt{s_{NN}}$, GeV\\\hline
$p+p^{\,\text a}$          & 22.0$^{\text b}$, 62.4, 200, 410$^{\text b}$, 500, 510 \\
$p+\mbox{Al}$              & 200 \\
$\mbox{O}+\mbox{O}$        & 200$^{\text b}$ \\
$p^{\,\text a}+\mbox{Au}$  & 200 \\
$\mbox{d}+\mbox{Au}$       & 19.6, 39.0, 62.4, 200 \\
$\mbox{He}^{3}+\mbox{Au}$  & 200 \\
$\mbox{Al}+\mbox{Au}$      & 4.9$^{\text b,c}$,200 \\
$\mbox{Cu}+\mbox{Cu}$      & 22.4$^{\text b}$, 62.4, 200 \\
$\mbox{Cu}+\mbox{Au}$      & 200 \\
$\mbox{Zr}+\mbox{Zr}$, $\mbox{Ru}+\mbox{Ru}^{\text d}$   & 200 \\
$\mbox{Au}+\mbox{Au}$      & FXT$^{\text c}$: 3.0, 3.2, 3.5, 3.9, 4.5, 5.2, 6.2, 7.2, 7.7 \\
                           & 7.7, 9.2$^{\text b}$, 11.5, 14.6, 17.3, 19.6, 27.0, 39.0, 54.4$^{\text b}$, 55.8$^{\text b}$, 62.4, 130, 200 \\
$\mbox{U}+\mbox{U}$        & 193 \\\hline
\multicolumn{2}{l}{\footnotesize $^{\text a}$ with unpolarized
($\sqrt{s}=62.4$ GeV) and with longitudinal / transverse polarized
beams;~ $^{\text b}$ run with}\\
\multicolumn{2}{l}{\footnotesize small integral luminosity;~
$^{\text c}$ run for STAR fixed target mode;~ $^{\text d}$ run
with isobar beams}
\end{tabular}
\end{table}

\begin{table}
\centering \caption{Data samples obtained during LHC runs.}
\label{tab-2}
\begin{tabular}{ll}\hline
Species & $\sqrt{s_{NN}}$, GeV\\\hline
$p+p$                 & 900, 2360, 2760, 5020, 7000, 8000, 13000, 13600 \\
$p+\mbox{Pb}$         & 5020, 8160 \\
$\mbox{Xe}+\mbox{Xe}$ & 5440 \\
$\mbox{Pb}+\mbox{Pb}$ & 2760, 5020, 5360 \\
\hline
\end{tabular}
\end{table}

\section{Some results for collectivity}\label{sec:2}

The section contains separate important results obtained within
very extensive studies of various collective effects which are
described in details in some recent reviews
\cite{arXiv-2303.17254-2023,EPJC-84-813-2024,arXiv-2404.06829-2024,arXiv-2405.10785-2024,arXiv-2405.18661-2024}.

\subsection{Hadronic jets}\label{sec:2.1}

The development of experimental technique and data analysis
results in the possibility for full reconstruction of jets created
not only in hadronic, mostly (anti)proton--proton ($p+p$,
$\bar{p}+p$) beam interactions, but also in nucleus--nucleus
($A+B$) collisions.

In $p+p$ the measurements for inclusive jet and dijet cross
sections are extended to the multi-TeV region for kinematic
parameters, namely, up to jet transverse momentum
$p_{T}^{\,\mbox{\scriptsize{J}}} \approx 4$ TeV/$c$ and dijet mass
$m_{\mbox{\scriptsize{JJ}}} \approx 10$ TeV/$c^{2}$
\cite{arXiv-2404.06829-2024}. Overall, fair agreement between the
measured cross sections (that span several orders of magnitude)
and the fixed-order pQCD calculations in next-to-next-to-leading
order (N$^{2}$LO), corrected for non-perturbative and electroweak
effects, is observed. The evolution of the strong coupling as a
function of the energy scale, $\alpha_{S}(Q)$ has been tested with
help of the ratio observable $R_{\Delta
\phi}(p_{T}^{\,\mbox{\scriptsize{J}}})$, related to the azimuthal
correlations among jets up to $Q \approx 2$ TeV/$c$, a higher
scale than that probed in previous measurements
\cite{arXiv-2405.18661-2024}. The $\alpha_{S}(m_{Z})$ has been
recently determined with help of the energy correlators inside
jets using an event sample of $\sqrt{s}=13$ TeV $p+p$ collisions.
The measured distributions are consistent with the trends in the
simulation that reveal two key features of the strong interaction:
confinement and asymptotic freedom, the strong coupling is
$\alpha_{S}(m_{Z})=0.1229_{-0.0050}^{+0.0040}$
\cite{PRL-133-071903-2024}. This result is the most precise
$\alpha_{S}(m_{Z})$ value obtained using jet substructure
observables and it is consistent with the world average
$[\alpha_{S}(m_{Z})]_{\mbox{\scriptsize{wa}}}=0.1180 \pm 0.0009$
\cite{PRD-110-030001-2024}. The pattern of the parton shower is
expected to depend on the mass of the initiating parton, through a
phenomenon known as the dead-cone effect, which predicts a
suppression of the gluon spectrum emitted by a heavy quark of mass
$m_{Q}$ and energy $E_{Q}$, within a cone of angular size
$m_{Q}/E_{Q}$ around the emitter. The QCD dead cone was directly
measured in $p+p$ collisions at $\sqrt{s}=13$ TeV with the
significances of $7.7\sigma$, $3.5\sigma$ and $1.0\sigma$ for $5 <
E_{Q} < 10$ GeV, $10 < E_{Q} < 20$ GeV and $20 < E_{Q} < 35$ GeV,
using iterative declustering of jets tagged with a fully
reconstructed $D^{0}$-meson \cite{EPJC-84-813-2024}. These results
provide direct sensitivity to the mass of quasi-free $c$ quarks,
before they bind into hadrons and pave the way for a study of the
mass dependence of the dead-cone effect, by measuring the dead
cone of $b$-tagged jets with a reconstructed beauty hadrons. The
energy frontier provides many unique approaches and discovery
opportunities for physics beyond Standard Model (SM) -- BSM
physics. A study of jets in $p+p$ collisions is focused, in
particular, on the search for the BSM physics within Effective
Field Theory (EFT) approach for top quark ($t$) sector up to
ultra-high energies
\cite{Okorokov-JPCF-1690-012006-2020,Okorokov-PAN-86-742-2022}.
The sensitivity to dijet resonances in $p+p$ is intensively
explored regarding of the projects future collider with proton
beams at different energies up to $\sqrt{s} = 500$ TeV
\cite{arXiv-2209.13128-2022,arXiv-2202.03389-2022}. The discovery
mass reach of a proton--proton collider is
$m_{\mbox{\scriptsize{JJ}}} \approx \kappa \sqrt{s}$ with
$\kappa=0.5$ for strongly produced dijet resonances, $\kappa=0.25$
for weakly produced ones \cite{arXiv-2202.03389-2022}.

In \cite{Okorokov-Proc-HEPFT-2014-PD} it was pointed out that
behavior of Higgs boson ($h$) in the quark-gluon matter at finite
temperature ($T$) called also (strongly coupled) quark-gluon
plasma -- (s)QGP\footnote{It should be noted that accelerator
complexes, at least, commissioned at present allow the study of
strongly couled QGP. Therefore, within the paper the abbreviation
QGP is used if it is disscused the corresponding state of strongly
interacting matter in general. The abbreviation sQGP is used if it
is considered the state of quark-gluon matter available for
experimental study at present.} -- could be studied at future
multi-TeV colliders. At present $h$ and $t$ are widely considered
as important probes of a pre-equilibrium stages of space-time
evolution of QGP in domain of high and ultra-high collision
energies
\cite{Okorokov-JPCF-1690-012006-2020,Okorokov-PAN-86-742-2022}.
Datasets with large integrated luminosities obtained at the LHC
for $p+\mbox{Pb}$ at $\sqrt{s_{NN}}=8.16$ TeV and
$\mbox{Pb}+\mbox{Pb}$ at $\sqrt{s_{NN}}=5.02$ TeV result in the
reconstruction of $\bar{t}t$ events in the (a) dilepton
($l^{+}l^{-}$), (b) $l^{+}l^{-}+2\,\mbox{jets}$ and (c)
$l^{\pm}+4\,\mbox{jets}$ channels become accessible
experimentally, where $l^{\pm}=e^{\pm}, \mu^{\pm}$. For
$p+\mbox{Pb}$ at $\sqrt{s_{NN}}=8.16$ TeV the most precision
result is obtained by ATLAS. Combining both (b) and (c) channels,
the $\bar{t}t$ pair production cross section is measured to be
$\sigma_{\bar{t}t}^{p+\mbox{\scriptsize{Pb}}}=(58.1 \pm
2.0^{+4.8}_{-4.4})$ nb, where the first uncertainty is statistical
and second one is systematic \cite{JHEP-2411-101-2024}. Cross
sections measured by ATLAS and CMS \cite{PRL-119-242001-2017} are
found to be in good agreement each other and with SM predictions.
Precision of these measurements opens a new way to constraint
parton distribution functions (PDFs) in the high-$x$ region. For
$\mbox{Pb}+\mbox{Pb}$ at $\sqrt{s_{NN}}=5.02$ TeV the first
measurements by CMS are
$\sigma_{\bar{t}t}^{\mbox{\scriptsize{Pb}}+\mbox{\scriptsize{Pb}}}=(2.5^{+0.8}_{-0.7})$
$\mu$b and $(2.0^{+0.7}_{-0.6})$ $\mu$b utilizing the leptons only
and in the channel (b) with the $b$ quarks respectively
\cite{PRL-125-222001-2020}, ATLAS obtained the result
$\sigma_{\bar{t}t}^{\mbox{\scriptsize{Pb}}+\mbox{\scriptsize{Pb}}}=
(3.6^{+1.0+0.8}_{-0.9-0.5})$ $\mu$b  using the channel with $e\mu$
and at least two jets \cite{arXiv-2411.10186-2024}. The
corresponding values of
$\sigma_{\bar{t}t}^{\mbox{\scriptsize{Pb}}+\mbox{\scriptsize{Pb}}}$
obtained by ATLAS and CMS agree within uncertainties for channel
(b). For the first case experimental result is consistent with
theoretical predictions using a range of different nuclear PDFs.
On the other side, the values extracted by CMS are compatible
with, though somewhat lower than, the expectations from scaled
$p+p$ data and pQCD calculations. The observation of $\bar{t}t$
production consolidates the evidence of the existence of all quark
flavors in the pre-equilibrium stage of the QGP at very high
energy densities, similar to the conditions present in the early
Universe \cite{arXiv-2411.10186-2024}.

Jet quenching is considered as one of the most promising
signatures of formation of the QGP and sensitive probe for
transport properties of final-state matter. Experiments at the LHC
extend the measurements for jet quenching up to $p_{T} \approx 1$
TeV/$c$. The nuclear modification factor ($R_{AA}$) exhibits
larger suppression for jets than hadrons at the same $p_{T}$, with
the ALICE jet spectrum extending down to $p_{T}=60$ GeV/$c$. At
higher $p_{T}$ the ALICE and ATLAS jet data are consistent, and
show slowly increasing $R_{AA}$ with increasing $p_{T}$
\cite{EPJC-84-813-2024}. In $\mbox{Pb}+\mbox{Pb}$ collisions at
$\sqrt{s_{NN}}=5.02$ TeV the $\gamma$-- and $b$--tagged jets are
less suppressed than the inclusive jets at
$p_{T}^{\,\mbox{\scriptsize{J}}} < 0.2$ TeV/$c$, implying that the
energy loss depends on the color charge and possibly the mass of
the parton. Suppression measurements in the heavy-flavour sector
at intermediate $p_{T}$ for nuclear collisions indicate that $b$
quarks lose less energy than $c$ quarks. These measurements are
described by models that include mass-dependent elastic energy
loss and a reduction of gluon radiation off heavier quarks, i.e.
the QCD dead-cone. Several collaborations also study the shift in
$p_{T}$ needed to match the $A+A$ spectra to that of the binary
scaled $p+p$. Such analysis show that at LHC energies $(\Delta
p_{T})_{q~\mbox{\scriptsize{dominated}}} < (\Delta
p_{T})_{g~\mbox{\scriptsize{dominated}}}$ and $(\Delta
p_{T})_{\mbox{\scriptsize{RHIC}}} < (\Delta
p_{T})_{\mbox{\scriptsize{LHC}}}$. In particular, jet energy loss
has been measured by ALICE experiment for the semi-inclusive
distribution of jets recoiling from a hadron trigger to be $\Delta
p_{T}=(8 \pm 2)$ GeV/$c$ for central $\mbox{Pb}+\mbox{Pb}$
collisions. This value is larger than that determined from similar
analysis at RHIC, though the comparison currently has limited
significance \cite{EPJC-84-813-2024}.

\subsection{Collective flows and chiral effects}\label{sec:2.2}

The geometrical shape of the impact region of two colliding
subatomic particles can be quantified by the Fourier decomposition
of the invariant distribution of the final-state particles, which,
in general case, is \cite{IJMPE-22-1350041-2013,PAN-80-1133-2017}
\begin{equation}
\displaystyle E\frac{\textstyle d^{3}N_{\alpha}}{\textstyle
d\bf{p}}=\frac{\textstyle 1}{\textstyle 2\pi} \frac{\textstyle
d^{\,2}N_{\alpha}}{\textstyle
p_{T}dp_{T}dy}\biggl[1+2\sum\limits_{n=1}^{\infty}\left\{v_{n,\alpha}\cos
\left(n \Delta \phi_{\alpha}\right)+a_{n,\alpha}\sin\left (n
\Delta \phi_{\alpha} \right)\right\}\biggr]. \label{eq:3.3.1}
\end{equation}
Here $\Delta \phi_{\alpha} \equiv
\phi_{\alpha}-\Psi_{\mbox{\footnotesize{RP}}}$, $\phi_{\alpha}$ is
an azimuthal angle of particle with certain sign of electric
charge $\alpha$ ($\alpha = +,-$) under study,
$\Psi_{\mbox{\footnotesize{RP}}}$ -- azimuthal angle of reaction
plane, $v_{n,\alpha}$ -- collective flow of $n$-th order, the
parameters $a_{n,\alpha}$ describe the effect of possible
$\mathcal{P/CP}$ violation.

Identified light hadron flow measurements have been performed for
a variety of collision systems recently. At RHIC, these include
collisions of $\mbox{Y}+\mbox{Au}$ at $\sqrt{s_{NN}}=200$ GeV with
$\mbox{Y} \equiv p$, $d$, ${}^{3}\mbox{He}^{2+}$, $\mbox{Cu}$ and
$\mbox{U}+\mbox{U}$ at $\sqrt{s_{NN}}=193$ GeV, while at the LHC
these include $\mbox{Xe}+\mbox{Xe}$ at $\sqrt{s_{NN}}=5.44$ TeV.
They have also been explored for higher orders of anisotropic
flow, with examples from RHIC for $\mbox{Au}+\mbox{Au}$ collisions
at $\sqrt{s_{NN}}=200$ GeV, and $\mbox{Pb}+\mbox{Pb}$ collisions
at $\sqrt{s_{NN}}=5.02$ TeV. A hallmark of the hydrodynamic
response is the mass ordering observed for the $p_{T}$-dependence
of elliptic flow ($v_{2}$) of various hadron species in the light
flavor sector \cite{arXiv-2303.17254-2023}. Through detailed
studies of azimuthal anisotropy coefficients $v_{n \geq 2}$ the
CMS data impose stringent constraints on the allowed range of the
shear viscosity-to-entropy ratio $\tilde{\eta} /s =0.08-0.20$. This
reaffirms that the sQGP behaves like a ``nearly perfect liquid'',
exhibiting minimal frictional momentum dissipation. Hydrodynamic
calculations can also describe measurements of higher-order flow
coefficients (up to $v_{8}$), non-linear contributions to
higher-order flow coefficients, and symmetry plane correlations.
On the other hand, for reasons subject to much theoretical
attention, hydrodynamic predictions cannot describe measured
anisotropic flow coefficients at LHC energies in ultra central
$\mbox{Pb}+\mbox{Pb}$ collisions to the same degree of accuracy as
mid-central $\mbox{Pb}+\mbox{Pb}$ interactions. The $J/\psi$ $v_{2}$
measured in $\mbox{Pb}+\mbox{Pb}$ collisions at
$\sqrt{s_{NN}}=5.02$ TeV by ALICE increases with $p_{T}$ and
reaches about 0.1 around $p_{T}=5$ GeV/$c$. Such a large $v_{2}$
signal at low to intermediate $p_{T}$ can only be explained by the
dominance of regenerated $J/\psi$ mesons inheriting the elliptic
flow of the constituent charm quarks which likely have reached
local thermalization in the QGP \cite{arXiv-2303.17254-2023}. This
observation constitutes a proof of deconfinement, as it implies
that coloured partons can move freely over distances much larger
than the hadronic scale \cite{EPJC-84-813-2024}.

The $c$ quarks are formed at the earliest stages of the collision,
and therefore will have to overcome much larger magnetic fields
than charged particles. The asymmetry measured by STAR for $v_{1}$
shows no charge asymmetry within the uncertainties and ALICE
observes linear dependence with pseudorapidity ($\eta$) for the differences of the
charge-dependent $v_{1}$, denoted as $\Delta v_{1}$, in
mid-central $\mbox{Pb}+\mbox{Pb}$ collisions at
$\sqrt{s_{NN}}=5.02$ TeV for charged particles and $D^{0}$,
$\bar{D}^{0}$ mesons. The slope $d\Delta v_{1}/d\eta$, extracted
with a linear fit function, yields $(1.68 \pm 0.49 \pm 0.41)
\times 10^{-4}$ for charged hadrons with $p_{T}>0.2$ GeV/$c$ for
the 5--40\% centrality interval and $(4.9 \pm 1.7 \pm 0.6) \times
10^{-1}$ for $D^{0}$ mesons with $3 < p_{T} < 6$ GeV/$c$ in a
centrality interval of 10--40\%, resulting in a significance of
2.6$\sigma$ and 2.7$\sigma$ for having a positive value,
respectively. This measurement constitutes the first experimental
hint of the existence of the initial state electromagnetic fields
at the LHC \cite{EPJC-84-813-2024}. The differences in the
measured global polarisation of $\Lambda$ and $\bar{\Lambda}$
provide an upper limit for the magnitude of the magnetic field at
freeze-out of $5.7 \times 10^{12}$ T and $14.4 \times 10^{12}$ T
at a 95\% confidence level (CL) in $\mbox{Pb}+\mbox{Pb}$
collisions at $\sqrt{s_{NN}}=2.76$ and 5.02 TeV, respectively
\cite{EPJC-84-813-2024}.

Atomic nuclei manifest a variety of shapes. In
\cite{Nature-635-67-2024} the collective-flow-assisted nuclear
shape-imaging method was introduced and used for the study of the
shape of ground-state ${}^{238}\mbox{U}^{92+}$ nuclei. It was
found a large deformation with a slight deviation from axial
symmetry in the nuclear ground state, aligning broadly with
previous low energy experiments \cite{Nature-635-67-2024}.

The strong evidence was found for collectivity through
multiparticle correlation analyses in $p+p$ and $p+\mbox{Pb}$
collisions at the LHC energies. Studies of particle correlation
functions have been extended by CMS to the even smaller
$\gamma+\mbox{Pb}$ collision system, using $p+\mbox{Pb}$ UPCs. The
$\gamma+\mbox{Pb}$ data are consistent with predictions of models
that do not include any collective effects. Measurements of prompt
$D^{0}$ and $J/\psi$ mesons in $p+p$ and $p+\mbox{Pb}$ collisions
at the LHC energies suggest a weaker collectivity signal for $c$
quarks than for light quarks. The study of $c$ quark collectivity
indicates positive $v_{2}$, even in low-multiplicity $p+\mbox{Pb}$
collisions, while heavier bottom hadrons are found to have weaker
collective signals, at a level that is currently not detected
conclusively, than those of light flavor hadrons
\cite{arXiv-2405.10785-2024}. The presence of a jet was shown to
alter the $v_{n}(p_{T})$ in the range of $p_{T} \sim 2-10$ GeV in
$p+\mbox{Pb}$ collisions. Thus, collective flow observables in
this $p_{T}$ range in small systems provide an intriguing
possibility to better understand jet-medium interactions
\cite{arXiv-2303.17254-2023}.

In a system that is not invariant under a parity transformation
(i.e., chiral), a electromagnetic current and electric dipole
moment of QCD matter can be induced by an external magnetic field,
such as generated in the passage of two (heavy) nuclei. The
resulting charge separation can be identified by studying the
$\mathcal{P/CP}$-odd sine terms in (\ref{eq:3.3.1}). The
experimental manifestation of the local topologically induced
$\mathcal{P/CP}$ violation in strong interaction is phenomenon
called chiral magnetic effect (CME). The correlator used for study
of CME can be generalised according to \cite{EPJC-84-813-2024}:
$\displaystyle \gamma_{m,n}=\langle
\cos[m\phi_{\alpha}+n\phi_{\beta}-(m+n)\Psi_{|m+n|}]\rangle$,
$\delta_{m}=\langle \cos[m(\phi_{\alpha}-\phi_{\beta})]$, $m,n \in
Z$ and $\Psi_{|m+n|}$ is the azimuthal angle of the symmetry plane
of $|m+n|$-th order. The presence of a net positive electric
charge can induce a positive axial current along the direction of
the magnetic field i.e., leading to flow of chirality. This is
caused by the chiral separation effect (CSE)
\cite{PAN-80-1133-2017}. The coupling between the CME and the CSE
leads to a wave propagation of the electric charge, resulting in
an electric charge quadrupole moment of the system, the chiral
magnetic wave (CMW) \cite{PAN-80-1133-2017}. The azimuthal
distribution of charged particles due to the presence of the CMW
can be written as $\displaystyle dN^{\pm} /
d\phi=N^{\pm}\{1+(2v_{2}\mp rA)\cos[2(\phi-\Psi_{2})]\}$, where
$A=(N^{+}-N^{-})/(N^{+}+N^{-})$ is the charge asymmetry, and $r$
is the parameter that encodes the strength of the electric
quadrupole due to the CMW \cite{EPJC-84-813-2024}. Therefore, one
can probe the value of $r$ by measuring the $v_{2}$ values for
different charges as a function of the charge asymmetry. Instead,
it was suggested
 to measure the covariance of $v_{n}$ and $A$ that is a robust observable
and does not depend on detector inefficiencies.

Isobaric collisions were proposed to study two systems with
similar $v_{2}$ but different magnetic field strengths, such as
${}^{96}\mbox{Ru}^{44+}$ and ${}^{96}\mbox{Zr}^{40+}$. The STAR
extracted an upper limit of the CME fraction of approximately 10\%
at a 95\% CL in isobar collisions at $\sqrt{s_{NN}}=200$ GeV
\cite{PRR-6-L032005-2024,PRC-110-014905-2024}. On the other hand,
the $\mbox{Au}+\mbox{Au}$ collision data from STAR indicate a
possible finite CME signal \cite{PRL-128-092301-2022}. This is
consistent with the expectation that the signal-to-background
ratio is approximately a factor of three larger in
$\mbox{Au}+\mbox{Au}$ collisions than in isobar collisions.
Measurements of $\Delta
\gamma_{1,1}=\gamma_{1,1}^{\mbox{\scriptsize{OS}}}-\gamma_{1,1}^{\mbox{\scriptsize{SS}}}$
at RHIC and LHC energies are qualitatively consistent with the CME
expectation, where $\gamma_{1,1}^{\mbox{\scriptsize{OS/SS}}}$
denotes the $\gamma_{1,1}$ with opposite (OS) / same (SS) sign
particle pair and the quantity $\Delta \gamma_{1,1}$ assumes
reduce of mutual backgrounds
\cite{IJMPE-22-1350041-2013,PAN-80-1133-2017}. One of the
difficulties in interpreting the positive $\Delta \gamma_{1,1}$ is
whether the CME is the major charge-dependent background
contribution to the observable, such as those from resonance
decays and jets \cite{NST-35-214-2024}. Using the event shape
engineering (ESE) method, the ALICE experiment showed that the CME
fraction in the measured $\Delta \gamma_{1,1}$ is consistent with
zero in $\mbox{Pb}+\mbox{Pb}$ collisions at $\sqrt{s_{NN}}=2.76$
TeV. The contribution from the CME to the measurement of charge
dependent correlations relative to the second order symmetry plane
($\gamma_{1,1}$) is constrained to an upper limit of 15--33\% in
$\mbox{Pb}+\mbox{Pb}$ and 2\% in $\mbox{Xe}+\mbox{Xe}$ collisions
at the LHC energies at 95\% CL \cite{EPJC-84-813-2024}. The
combination of the second and third harmonic results for CMW
studies at the LHC indicates a significant background contribution
from local charge conservation \cite{EPJC-84-813-2024}. The CMS
results obtained for small systems unambiguously demonstrate that
the CME and CMW signals in nuclear collisions are too small to be
observed at the LHC energies. The most stringent upper limit to
date has been set on the CME signal which is estimated to be 13\%
in $p+\mbox{Pb}$ at $\sqrt{s_{NN}}=8.16$ TeV and 7\% in
$\mbox{Pb}+\mbox{Pb}$ collisions at $\sqrt{s_{NN}}=5.02$ TeV, at
95\% CL \cite{arXiv-2405.10785-2024}.

\subsection{Femtoscopic correlations and Bose--Einstein
condensation}\label{sec:2.3}

The correlations at low relative momentum, called also as
femtoscopic correlations, emerge due to both the symmetrization
requirement of quantum statistics (QS) and the effect of final
state interaction (FSI) among particles of the system under
consideration. Experimentally, correlation function (CF) in
general case of $n$-particle system is defined as the ratio
$\displaystyle C_{n}=\zeta \mathcal{N}_{\mbox{\scriptsize{s}}} /
\mathcal{N}_{\mbox{\scriptsize{m}}}$, where the quantities are the
functions of the set $\{p_{i}\}_{i=1}^{n} \equiv
(p_{1},\dots,p_{n})$ of 4-momenta $p_{i}$, $i=1,\dots,n$ for
secondary particles, $\mathcal{N}_{\mbox{\scriptsize{s}}}$ and
$\mathcal{N}_{\mbox{\scriptsize{m}}}$ represent the distributions
for particles produced in the same and in different collisions,
respectively, $\zeta$ denotes the corrections for all experimental
effects (acceptance, particle identification etc.). At present the
correlations of identical particles with low relative momenta are
mostly used for study of space-time extents of fireball. The main
part of experimental data are obtained for pairs of $\pi^{\pm}$,
also there are some results for $pp$ ($\bar{p}\bar{p}$), neutral
particle ($\gamma$, $\pi^{0}$, $K^{0}$, $\Lambda$) and charged
kaon pairs. The FSI effect allows for the correlation femtoscopy
with unlike particles the access, in particular, to a study of
strong interactions between specific particles. This issue is in
the focus below.

The strong potentials predicted for the four allowed spin and
isospin states of the $p-\Xi^{-}$ system can be found elsewhere
\cite{ARNPS-71-377-2021}. It was obtained that for all cases, an
attractive interaction and a repulsive core characterize the
potentials. The total $p-\Xi^{-}$ CF measured in $p+p$ collisions
at $\sqrt{s}=13$ TeV recorded with a high-multiplicity trigger by
ALICE lies above the Coulomb predictions as well as the CF for
$p-\Omega+\bar{p}-\bar{\Omega}$ \cite{EPJC-84-813-2024},
demonstrating the presence of an additional attractive strong
interaction between $p$ and multistrange hyperons. The
$p-\Omega+\bar{p}-\bar{\Omega}$ CF measured by STAR in
$\mbox{Au}+\mbox{Au}$ collisions at $\sqrt{s_{NN}}=200$ GeV does
not allow to extract the interaction parameters due to limited
statistics. However, based on the comparison of experimental
results and model predictions for CF, one can conclude that data
obtained by STAR favor a positive scattering length for the
$p-\Omega$ interaction \cite{PLB-790-490-2019}. The positive
scattering length and the measured ratio of the
$p-\Omega+\bar{p}-\bar{\Omega}$ CFs from peripheral to central
collisions less than unity for $k^{*} < 40$ MeV/$c$ within
$1\sigma$ favors the $p-\Omega$ interaction potential with deep
$p-\Omega$ bound state and binding energy $E_{b} \sim 27$ MeV for
$p$ and $\Omega$, where $k^{*}=|{\bf p}^{*}_{1}-{\bf
p}^{*}_{2}|/2$ is the absolute value of relative 3-momentum of one
of the particles in the pair rest frame. Baryon-antibaryon
correlations were studied in $p+p$ and $\mbox{Pb}+\mbox{Pb}$
collisions at the LHC and the real and imaginary part of the
scattering parameters for $p-\Lambda$ and $\Lambda-\bar{\Lambda}$
were extracted for the first time \cite{EPJC-84-813-2024}.

Knowledge regarding the interaction of hyperons with nucleons is
one of the key ingredients needed to understand the composition of
the compact astrophysical objects, in particular, neutron stars
(NSs). The high-density environment ($\rho \approx 3-4\rho_{0}$)
that is supposed to occur in the interior of NSs leads to an
increase in the Fermi energy of the nucleons, translating into the
appearance of new degrees of freedom, such as hyperons. The
inclusion of hyperons leads to NS configurations that cannot reach
the current highest mass limit from experimental observations of
$2.2M_{\odot}$. For this reason, the presence of hyperons inside
the inner cores of NSs is still under debate, and this so-called
hyperon puzzle is far from being solved \cite{ARNPS-71-377-2021}.
A major advance in understanding the role played by heavier
strange hadrons in the hyperon puzzle has been achieved by the
validation of lattice QCD predictions for the $N\Xi$ interaction
by the aforementioned experimental results for $p-\Xi$
correlations. Mass-radius relationship $M(R)$ for the EoS scenario
taking into account the constraints from the recent accelerator
data for $p-\Xi$ interaction agree reasonably with the
astrophysical measurements of heavy NSs \cite{ARNPS-71-377-2021}.
In particular, this EoS allows the existence of heavy NS with $M >
2M_{\odot}$ at radii within the narrow range 11.5--12.5 km, which
is fully compatible with the recent measurements of NSs close to
and above two solar masses \cite{ARNPS-71-377-2021}.

System with arbitrary number of bosons can undergo a
Bose--Einstein condensation (BEC) due to statistical properties of
quantum system and symmetry of the wave function (WF) of a boson
state. Regarding the multiparticle production process the
increasing the number density of bosons or increasing the overlap
of the multi-boson wave-packet states, achieved by changing the
size of the single-particle wave-packets lead to condensation of
bosons into the same quantum state and bosonic (pion) laser could
be created. The first case is the pion laser model (PLM)
\cite{PLB-301-159-1993} and second approach is called generalized
pion laser model (gPLM) \cite{PRL-80-916-1998,HIP-9-241-1999}.

The coherent emission can be considered as one of the experimental
signals of the appearance of BEC. Search for coherent pion
emission was studied with help of multiparticle correlations
within femtoscopy in $p+p$, $p+\mbox{Pb}$ and
$\mbox{Pb}+\mbox{Pb}$ at the LHC energies
\cite{PRC-93-054908-2016}. The measured same-charge multipion
correlations are compared to the expectation from lower-order
experimental CFs. There is no a significant suppression of 4-pion
correlations in $p+p$ or $p+\mbox{Pb}$ collisions, although the
unknown strength of the nonfemtoscopic background prevents an
absolute statement. A significant suppression of multipion
Bose--Einstein correlations has been observed in
$\mbox{Pb}+\mbox{Pb}$ collisions at $\sqrt{s_{NN}}=2.76$ TeV. A
coherent fraction of about $0.32 \pm 0.03 \pm 0.09$ could largely
explain the 4 pion suppression, but the same value cannot explain
the 3 pion suppression. Thus the origin of the suppression is not
clear and the effect may be explained by postulating either
coherent pion emission or large multibody Coulomb effects
\cite{PRC-93-054908-2016}.

Within phenomenological studies of BEC in strong interaction
processes
\cite{Okorokov-AHEP-2016-5972709-2016,Okorokov-PAN-82-838-2019,Okorokov-PAN-87-172-2024}
the particle density is defined as follows: $ \displaystyle
n_{\scriptsize{\mbox{ch}}}=N_{\scriptsize{\mbox{ch}}} / V$, where
$N_{\scriptsize{\mbox{ch}}}$ is the total charged particle
multiplicity, $V$ -- estimation for the volume of the emission
region of the boson under consideration (pions). The physical
quantities in r.h.s. of the above equation --
$N_{\scriptsize{\mbox{ch}}}$ and $V$ -- are model-dependent. The
critical value for $n_{\scriptsize{\mbox{ch}}}$ can be calculated
with help of the equation above and transition to the critical
total multiplicity. The following relation
\begin{eqnarray}
\displaystyle
N_{\scriptsize{\mbox{ch,c}}}&=&\frac{1}{N_{\pi,(1)}}\biggl[\frac{1+X+\sqrt{1+2X}}{2}\biggr]^{3/2}, \\
X& \equiv&
2m_{\pi}T_{\scriptsize{\mbox{eff}}}R_{\scriptsize{\mbox{eff}}}^{2},~
T_{\scriptsize{\mbox{eff}}}=T+\frac{\Delta_{p}^{2}}{2m_{\pi}},~
R_{\scriptsize{\mbox{eff}}}^{2}=R_{m}^{2}+\frac{T}{2\Delta_{p}^{2}
T_{\scriptsize{\mbox{eff}}}}.\nonumber \label{eq:3.12}
\end{eqnarray}
is suggested elsewhere \cite{Okorokov-PAN-87-172-2024} for the
critical value of $N_{\scriptsize{\mbox{ch}}}$ for 3D case based
on the gPLM. Here $T_{\scriptsize{\mbox{eff}}}$ and
$R_{\scriptsize{\mbox{eff}}}$ is effective temperature and radius
of the source, $\Delta_{p}$ is the momentum spread of the emitted
pions, $N_{\pi,(1)}=0.25$ is the fraction of the 1-st generation
pions to be emitted from a static Gaussian source within unit of
$\eta$, $R_{m}$ is the estimation of the source radius, $T \approx
T_{\scriptsize{\mbox{ch}}}$ is the source temperature supposed
equal to the value of the parameter at chemical freeze-out
\cite{Okorokov-AHEP-2016-5972709-2016,Okorokov-PAN-82-838-2019,Okorokov-PAN-87-172-2024}.
The energy-dependent average $N_{\scriptsize{\mbox{ch}}}$ in $p+p$
collisions is approximated by functions $\displaystyle \langle
N_{\scriptsize{\mbox{ch}}}^{pp}\rangle_{1} \propto
\varepsilon^{0.29}$ and $\langle
N_{\scriptsize{\mbox{ch}}}^{pp}\rangle_{2}=\langle
N_{\scriptsize{\mbox{ch,F}}}\rangle+N_{0}$ described in details
elsewhere \cite{Okorokov-PAN-87-172-2024}, where $\langle
N_{\scriptsize{\mbox{ch}}}^{pp}\rangle_{2}$ was inspired by pQCD
and $\varepsilon \equiv s/s_{0}$, $s_{0}=1$ GeV$^{2}$. For $A+A$
collisions the functions $\displaystyle \langle
N_{\scriptsize{\mbox{ch}}}^{AA}\rangle_{1} \propto
\varepsilon^{0.55}_{NN}$ and $ \displaystyle \langle
N_{\scriptsize{\mbox{ch}}}^{AA}\rangle_{2} \propto
\varepsilon^{0.15}_{NN}\ln\varepsilon_{NN}$ are recently used
\cite{Okorokov-PAN-87-172-2024}. In the case of the Poissonian
distribution with mean $n_{0}$ for the multiplicity of secondary
bosons the influence of the BEC results in the modified
probability distribution for the special case of the rare Bose gas
\cite{HIP-9-241-1999}, i.e. $X \gg 1$, with the mean value
$\displaystyle n(X)=n_{0}[1+n_{0}(2X)^{-3/2}]$. Phenomenological
studies of possible manifestation of BEC in various strong
interaction processes
\cite{Okorokov-AHEP-2016-5972709-2016,Okorokov-PAN-82-838-2019,Okorokov-PAN-87-172-2024}
prove that $\langle n_{\scriptsize{\mbox{ch}}}\rangle$  in $p+p$
is smaller than its critical value up to $\sqrt{s} \sim 1$ PeV for
any used views of $\langle
N_{\scriptsize{\mbox{ch}}}^{pp}\rangle$. In heavy ion $A+A$
collisions $n_{\scriptsize{\mbox{ch}}}$ larger than its critical
value at any energies $10^{17} \leq E_{N} \leq 10^{21}$ eV in
laboratory reference system (l.r.s.) for any approximations of
$\langle N_{\scriptsize{\mbox{ch}}}^{AA}\rangle$ considered in
\cite{Okorokov-PAN-82-838-2019,Okorokov-PAN-87-172-2024}. The BEC
results in to the visible increase of charged particle density at
even large enough $X=5$ for the energy range with $\langle
n_{\scriptsize{\mbox{ch}}}^{AA}\rangle > \langle
n_{\scriptsize{\mbox{ch,c}}}^{AA}\rangle$. The absence of clear
manifestation of BEC in $\mbox{Pb}+\mbox{Pb}$ collisions at
$\sqrt{s_{NN}}=2.76$ TeV well agrees with the conclusion within
gPLM \cite{Okorokov-PAN-87-172-2024}. The characteristics
\begin{equation}
\displaystyle z_{\pi}^{(n)}=\frac{\ln \langle
n_{\scriptsize{\mbox{ch,BEC}}}^{AA}\rangle-\ln \langle
n_{\scriptsize{\mbox{ch,0}}}^{pp}\rangle}{\ln \langle
n_{\scriptsize{\mbox{ch,0}}}^{AA}\rangle-\ln \langle
n_{\scriptsize{\mbox{ch,0}}}^{pp}\rangle},~~~ \Delta
z_{\pi}^{(n)}=z_{\pi}^{(n)}-1. \label{eq:3.18}
\end{equation}
determined in \cite{Okorokov-PAN-87-172-2024} are used for
quantitative study of the effect of BEC on the density of
secondary charged pions. Here $\langle
n_{\scriptsize{\mbox{ch,BEC}}}^{AA / pp}\rangle$ is the average
density of charged pions with taking into account the possible BEC
at the region of (kinematic) parameter space with $\langle
n_{\scriptsize{\mbox{ch}}}\rangle > n_{\scriptsize{\mbox{ch,c}}}$
in $A+A$ or $p+p$ collisions respectively, $\langle
n_{\scriptsize{\mbox{ch,0}}}^{AA / pp}\rangle$ is the average
particle density when the BEC is switched off in the fixed type
interaction\footnote{The superindex ``(n)'' means that the
quantities are calculated with help of the average densities
namely but not via the average total multiplicities.}.

\begin{figure*}
\includegraphics[width=14.0cm,height=14.0cm]{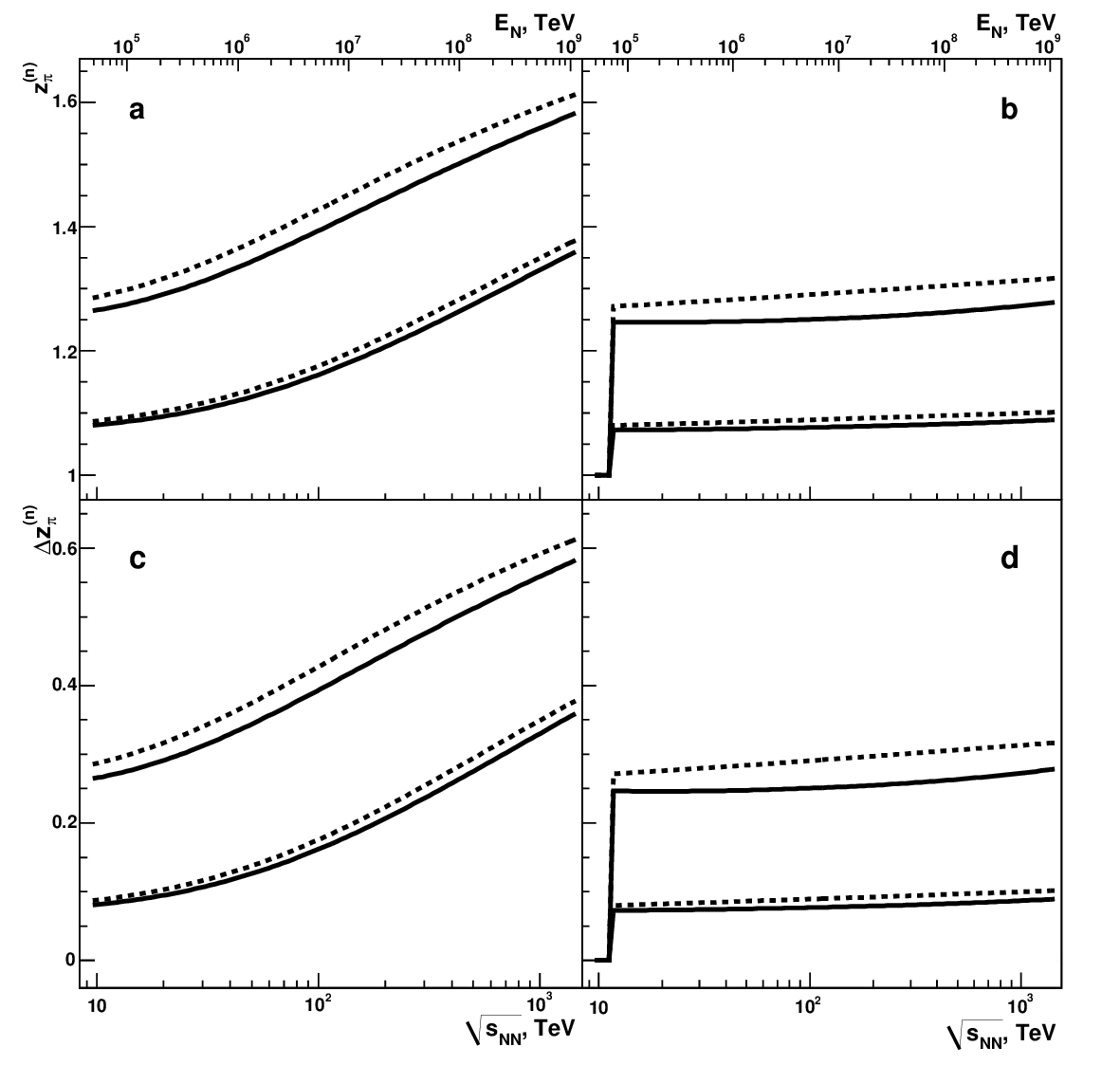}
\vspace*{8pt} \caption{Parameters $z_{\pi}^{(n)}$ (a, b) and
$\Delta z_{\pi}^{(n)}$ (c, d) in dependence on energy. In the case
of a symmetric ($A+A$) ion collisions the approximation $\langle
N_{\scriptsize{\mbox{ch}}}^{AA}\rangle_{1}$ is used for the panels
(a, c) while $\langle N_{\scriptsize{\mbox{ch}}}^{AA}\rangle_{2}$
is used for the panels (b, d). In each panel solid lines
correspond to the $\langle
N_{\scriptsize{\mbox{ch}}}^{pp}\rangle_{1}$, dashed lines are for
$\langle N_{\scriptsize{\mbox{ch}}}^{pp}\rangle_{2}$. Effect of
BEC is taken into account in accordance with the relation for the
mean value $n(X)$ shown in the main text for energy region with
$\langle n_{\scriptsize{\mbox{ch}}}\rangle >
n_{\scriptsize{\mbox{ch,c}}}$ in certain type of collisions if
any. The upper collection of curves are for $X=2$, lower curves
are for $X=5$ \cite{Okorokov-PAN-87-172-2024}.}\label{fig01}
\end{figure*}

Fig.~\ref{fig01} shows energy dependence of $z_{\pi}^{(n)}$ (a, b)
and $\Delta z_{\pi}^{(n)}$ (c, d) for all possible combinations of
approximations $\langle
N_{\scriptsize{\mbox{ch}}}^{pp}\rangle_{1,2}$ and $\langle
N_{\scriptsize{\mbox{ch}}}^{AA}\rangle_{1,2}$ in $p+p$ / $A+A$
collisions. In general, Figs. \ref{fig01}a, b demonstrate that the
curves for $z_{\pi}^{(n)}$ show the close behavior for various
$\langle N_{\scriptsize{\mbox{ch}}}^{pp}\rangle$, especially at
larger $X$. The clear increase of $z_{\pi}^{(n)}$ is observed with
growth of energy in the case of the approximation $\langle
N_{\scriptsize{\mbox{ch}}}^{AA}\rangle_{1}$ (Fig. \ref{fig01}a)
whereas there is almost no dependence $z_{\pi}^{(n)}$ vs energy
for $\langle N_{\scriptsize{\mbox{ch}}}^{AA}\rangle_{2}$
especially at $X=5$ in the domain with the presence of BEC effect
(Figs. \ref{fig01}b). Values of $z_{\pi}^{(n)}$ are noticeably
larger for calculations at $X=2$ with equation $\langle
N_{\scriptsize{\mbox{ch}}}^{pp}\rangle_{2}$ than that for the
approximation $\langle N_{\scriptsize{\mbox{ch}}}^{pp}\rangle_{1}$
in any considered cases of parameterization for $\langle
N_{\scriptsize{\mbox{ch}}}^{AA}\rangle$ vs energy. This
discrepancy is some clearer for the function $\langle
N_{\scriptsize{\mbox{ch}}}^{AA}\rangle_{2}$ in the domain $E_{N}
\gtrsim 10^{19}$ eV (Fig. \ref{fig01}b). As expected, the features
of the behavior of $\Delta z_{\pi}^{(n)}$ in dependence on energy
parameters (Figs. \ref{fig01}c, d) are the same as well as the
above observations for $z_{\pi}^{(n)}$ due to relation
(\ref{eq:3.18}). Thus within the special case of the gPLM the both
parameters $z_{\pi}^{(n)}$ and $\Delta z_{\pi}^{(n)}$ show the
increase of pion yield for the case of presence of BEC and
magnitude of this increase does not contradict, at least, at
qualitative level to the muon excess observed in the collisions of
ultra-high energy cosmic ray (UHECR) particles
\cite{Okorokov-PAN-87-172-2024}.

\section{Summary}\label{sec:3}

The research programs of the large experiments at RHIC and the LHC
open a multimessenger era for the physics of strong interaction.

In $p+p$ collisions at the LHC jet production is studied up to
TeV-region in jet $p_{T}$ and virtuality scale $Q$. Measurements
of jet cross sections and strong coupling $\alpha_{S}$ are well
agree with QCD calculations. Direct observation of dead cone
effect supports the feature of the evolution of parton shower for
heavy quarks predicted by QCD. The difference observed between
ridge yield in low-multiplicity $p+p$ events at the LHC and in
$e^{+}e^{-}$ annihilation at LEP energies indicates on the
additional processes for particle production in $p+p$
interactions. Enormous efforts on theory and experiment for
nuclear collisions during last decades resulted in, with strong
help of the study of collective effects, to the ``Standard model''
of (heavy) ion interactions at high energies. The cross sections
are measured for top pair production in $p+\mbox{Pb}$ and
$\mbox{Pb}+\mbox{Pb}$ collisions in multi-TeV energy domain. A
global Bayesian analysis of jet and jet substructure data from
RHIC and the LHC allows the extraction of the sQGP transport
coefficient $\hat{q} / T^{3}$ which smooth decrease from on about
6.5 to 4.0 at increase of $T$ from 0.15 to 0.50 GeV.

A key development during recent time was the democratization of
initial state models that can reproduce experimental data for
azimuthal anisotropy coefficients. The study of elliptic flow
allows the new method which images the nuclear global shape by
colliding them at ultra relativistic speeds and analysing the
collective response of outgoing debris. The highest ever values of
anisotropic and radial flow in heavy-ion collisions are achieved
at the LHC with radial-flow velocities up to about $0.7c$ and
light hadron $v_{2}$ measurements, which determine the magnitude
of elliptic flow, are 30\% higher than at the top RHIC energy.
Measurements in collisions of small systems have found signatures
of sQGP similar to those observed in large systems. Extensive and
various studies allow the indication on the CME and CMW signals in
$\mbox{Au}+\mbox{Au}$ collisions with the upper limit for the
first case $\sim 10\%$ from isobar $\mbox{Ru}+\mbox{Ru}$ and
$\mbox{Zr}+\mbox{Zr}$ collisions. Direct studies for the existence
of CME and CMW in heavy-ion collisions revealed that background
effects are dominating at the LHC with the upper limit for the
first case $\sim 20\%$ in $\mbox{Pb}+\mbox{Pb}$ collisions at
$\sqrt{s_{NN}}=5.02$ TeV.

At present femtoscopic correlations are actively studied for very
wide set of particle species including charmed mesons and light
nucleus. Taking into account the femtoscopic results for $p-\Xi$
lead to mass--radius dependence for NS which agree with
astrophysical data. Possibly, the BEC may affect on soft pion
production in, at least, heavy nucleus collisions in multi-TeV
range of $\sqrt{s_{NN}}$. This phenomenon provides noticeable
increase the mean values of particle density as well as total
multiplicity of charged particles (pions). Aforementioned new
feature of multiparticle processes can in the general case
contribute to the muon yield recorded in collisions of UHECR
particles with the atmosphere.

\newpage

\end{document}